\documentclass[12pt,twoside]{article}
\usepackage{fleqn,espcrc1}

\usepackage{graphicx}
\usepackage[figuresright]{rotating}

\newcommand{\AmS}{{\protect\the\textfont2
  A\kern-.1667em\lower.5ex\hbox{M}\kern-.125emS}}

\title{Nuclear Physics on the Light Front}

\author{Gerald A. Miller \address{Department of Physics, University of
  Washington,\\  
        Seattle, Washington 98195-1560 USA}
             \thanks{This work is partially supported by the USDOE. }} 
       
\begin{document}

\maketitle

\begin{abstract}
High energy scattering experiments involving nuclei are typically
analyzed in terms
of light front variables. The desire to provide realistic, relativistic
wave functions expressed in terms of these variables led me to try to
use  light front dynamics to compute nuclear wave functions.
The progress is summarized here.

\end{abstract}

\section{INTRODUCTION}

Much of nuclear physics is concerned with the transition between nucleon-meson
and quark-gluon degrees of freedom, with deep-inelastic lepton-nucleus
scattering, Drell-Yan production and high energy (e,e'p) and (p,pp) reactions
being some of the tools of investigation. Since high energies are involved,
it is absolutely necessary to incorporate relativity. Our project is to
implement Poincare invariance in nuclear physics by using light front dynamics.
This is useful in analyzing experiments because the canonical momentum
variable,
$k^+$,  is closely related to experimental observables. The challenge is
to use these variables while incorporating the dynamical richness necessary to
properly describe nuclei.
 Consider the plus-momentum  distributions $f_{B,M}(k^+)$ which give the
 probability that a nuclear
 baryon $B$ or meson $M$ has a plus-momentum of $k^+$.
 If the relevant nuclear wave
functions depend on $k^+$, the canonical spatial variable is
$x^-=x^0-x^3$. This leaves $x^+=x^0+x^3$ to be used as a time variable.
Thus the light front dynamics is very different than the usual dynamics.
If this light front formalism is used, $f_{B,M}$ are simply related to the
square of the ground state nuclear wave function. Thus the use of light front
dynamics provides vast simplicities, provided one can obtain the ground state
wave function using realistic Lagrangians and many-body techniques.
It is easier to carry out such a program for the effective
hadronic Lagrangians of
nuclear physics than for $QCD$ because hadrons all have a non-zero mass and
because the vacuum is not a condensed state of nucleon-anti-nucleon pairs.
Another advantage is that the Fock state expansion of the wave function is
in terms of on-shell particles.

The remainder is meant to present a brief outline of the progress. The main
goal is to provide a series of examples to show that the  light front
can be
used for high energy realistic-nuclear physics.
                                
\section{ LIGHT  FRONT  QUANTIZATION OF HADRONIC LAGRANGIANS}
You have to have a Lagrangian  ${\cal L}$   no matter how bad. We use
  ones in which the degrees of freedom are 
nucleons, vector and scalar mesons and pions. The  existence of ${\cal L}$ 
allows the  derivation of
the canonical-symmetric energy momentum tensor $T^{\mu\nu}$.
In light front dynamics the momentum is $P^+=P^0+P^3$ where $P^\mu$
is the total
momentum
operator: $P^\mu={1\over 2}\int d^2x_\perp dx^-\;T^{+\mu}$.
The $x^+$ development operator is $P^-$. One necessary detail is that
that $T^{+-}$ must be expressed in terms of independent degrees of freedom.
One uses the equations of motion to express the dependent degrees of freedom
in terms of the independent ones, and uses these constraint equations in the
expression for $T^{+-}$. Except for chiral Lagrangians,
the necessary quantization procedure
has been carried out long ago by
other authors\cite{des71}-\cite{yan34}.

\section{ INFINITE NUCLEAR MATTER IN MEAN FIELD APPROXIMATION: PLUS-MOMENTUM
 DISTRIBUTIONS}
The first problem we tried to solve is that of understanding nuclear matter
in the mean field approximation using the Walecka model. This problem has been
solved in a manifestly covariant manner\cite{bsjdw},
so that the existing results
for the energy of the system provide a useful check of our light front
procedure. We are able to reproduce the standard results
\cite{Miller:1997xh,Miller:1997cr}.

The interesting feature of light front dynamics that one is able to calculate
the plus-momentum distributions. For this simple problem we found
that the nuclear vector mesons
carry about a third of the nuclear plus momentum,  but
       their momentum distribution has support only at $k^+ =0$,
with\cite{Burkardt:1998bt} $k^+f_v(k^+)=0.35M_N\delta(k^+)$. Thus
 the vector mesons do not contribute to nuclear deep inelastic
       scattering.
       This zero mode effect occurs because in the mean field approximation,
       the meson fields of an infinite
       system   are constant in both space and time.

       This is an intriguing result which is caused by the large values of the
       scalar and vector potentials, which are characteristic of the Walecka
       model.  The nucleons carry only 65\% of the
       nuclear plus-momentum, a result in severe disagreement with deep
       inelastic scattering data. It is necessary to see if it survives for
       the case of finite-sized nuclei, and if correlations  between nucleons
       are
       included.

\section{FINITE NUCLEI IN MEAN FIELD APPROXIMATION}
 We showed\cite{Blunden:1999hy,Blunden:1999gq}
 that the necessary variational principle is a constrained one
       which
       fixes the expectation
       value of the total momentum operator $P^+$
       to be the same as that for $P^-$.
       This is the
       same as minimizing
       the sum of the total momentum operators: $P^-+P^+$.
       A new  light-front version of the
       equation for  the single nucleon modes was obtained, and a new numerical
       technique for their solution was introduced.
         The ground state wave function is treated as a
       meson-nucleon Fock state, and
       the meson
       fields are treated as expectation
       values of field operators in that ground
       state. The resulting equations
       for these expectation
       values was  shown to be closely related to the usual meson field
       equations.
       The computed binding energies  are
       essentially the same as for the usual equal-time theory.
       The nucleon plus momentum distribution
$f_N(k^+)$        
       peaks for  $k^+$ about seventy
       percent of the nucleon mass, which again is far too small to be
       consistent with deep inelastic scattering data.
       The mesonic component of the ground state
       wave function
       was used to
       determine the scalar and vector meson momentum distribution
       functions, and 
       the
       vector mesons were found to carry about thirty percent of the nuclear
       plus-momentum. We are currently investigating other Lagrangians
       which lead to smaller magnitudes of scalar and vector potentials and
       therefore yield better descriptions of the deep inelastic scattering
       data. 
\section{PIONS AND CHIRAL SYMMETRY }
Light front quantization of a chiral Lagrangian can be performed
\cite{Miller:1997cr}, if one
uses a Lagrangian due to G\"{u}rsey\cite{gur}.
Pion-nucleon scattering at tree level was shown to reproduce soft pion theorems
\section{NUCLEON-NUCLEON SCATTERING}

The Weinberg-type  equation\cite{fs}  is the light front version of the
Lippman-Schwinger equation. This equation is equivalent to the the
Blankenbecler-Sugar equation, except that retardation effects
need to be included\cite{Miller:1997cr}. Our hadronic chiral Lagrangian
was used\cite{Miller:1998tp,Miller:1999ap}
 to obtain a light front version
       of a one-boson-exchange nucleon-nucleon potential (OBEP).
       The accuracy of our description of the nucleon-nucleon
       (NN) data is good,
       and similar to that of other relativistic OBEP models. 
\section{ NUCLEAR MATTER WITH  NUCLEON-NUCLEON CORRELATIONS}

The trivial nature of the vacuum in the light front formalism was exploited in
deriving\cite{Miller:1998tp,Miller:1999ap}
the equations analogous to the 
       the Hartree-Fock and Brueckner Hartree-Fock equations.
       Applying our light front OBEP, the nuclear matter
       saturation properties are reasonably well reproduced. The computed
        value of the compressibility, 180 MeV,  is
       smaller than that of alternative relativistic approaches to nuclear
       matter in which the compressibility
       usually comes
       out too large.
      We   showed that replacing the
       meson degrees of freedom by a
       NN interaction is a consistent approximation, and
       that the formalism allows one to
       calculate corrections to this approximation
       in a well-organized manner.
        The mesonic Fock space components of the
       nuclear wave function are studied  also, and
       aspects of the meson and nucleon plus-momentum distribution functions
       are computed.
       We find that there are about 0.05 excess pions per nucleon.

The magnitudes of the scalar and vector potentials
are far smaller than found in the mean field approximation.
If we 
neglect the influence of
two-particle-two-hole states
to approximate
$f(k^+)$ 
the  nucleons are found to carry 81\% (as opposed to the 65\% 
of mean field theory) of  the nuclear plus momentum. 
This represents a vast improvement in the description of nuclear deep inelastic
scattering as the  
minimum value of the ratio $F_{2A}/F_{2N}$
 is increased by a factor of twenty   towards 
the data, which is not enough to provide a satisfactory description. 
We can be optimistic about future
results because including nucleons
with momentum greater than $k_F$ would substantially increase the
computed ratio $F_{2A}/F_{2N}$ since  $F_{2N}(x)$ decreases very rapidly
with increasing values of $x$ and because $M^*$ would increase at high momenta

Turn now to the experimental information about the nuclear pionic content.
The Drell-Yan experiment on nuclear targets \cite{Alde:1990im}
showed no enhancement of nuclear pions within an error of about 5\%-10\% for
their heaviest target. 
Understanding this result is 
an important challenge to the 
understanding of nuclear dynamics~\cite{missing}. 
Here we have a good description of nuclear dynamics, 
and our 5\%  enhancement is consistent\cite{jm},
within errors, with the Drell-Yan
data.

\section{SUMMARY} 

The light front approach has now been applied to infinite nuclear matter in the
mean field approximation, finite-sized  nuclei in the same approximation,
$\pi N$ and $NN$ scattering, and to correlated nucleons in
infinite nuclear matter 
Thus it seems that  one can use the light front approach to compute nuclear
energies, wave functions and the experimentally important plus-momentum
distributions. There are indications that the computed quantities will
ultimately be in good agreement with experiment. But
the use of  light front dynamics in nuclear physics is only in its infancy,
and much remains to be done to understand nuclear deep inelastic data the
future expected flood of data on the (e,e'p) and (p,pp) reactions.

\end{document}